\documentclass[journal]{IEEEtran}

\usepackage{xcolor,soul,framed} 

\colorlet{shadecolor}{yellow}
\usepackage[pdftex]{graphicx}
\graphicspath{{../pdf/}{../jpeg/}}
\DeclareGraphicsExtensions{.pdf,.jpeg,.png}

\usepackage[cmex10]{amsmath}
\usepackage{array}
\usepackage{mdwmath}
\usepackage{mdwtab}
\usepackage{eqparbox}
\usepackage{url}
 \usepackage[justification=centering]{caption}
 \usepackage{diagbox}
\usepackage{booktabs} 
\usepackage{caption}
\usepackage{amsfonts}
\usepackage{amsmath}
\usepackage{amssymb}

\begin{document}
\bstctlcite{IEEEexample:BSTcontrol}
    \title{Joint Radio Frequency Fingerprints Identification via Multi-antenna Receiver}
  \author{Xiaofang~Chen,~\IEEEmembership{Student Member,~IEEE,}
      Wenbo~Xu,~\IEEEmembership{Member,~IEEE,}
      and Yue~Wang,~\IEEEmembership{Senior Member,~IEEE}

  \thanks{XXX.}
  \thanks{XXX.}
  \thanks{XXX.}%
  \thanks{XXX.}
  \thanks{XXX.}}

\markboth{IEEE TRANSACTIONS ON MICROWAVE THEORY AND TECHNIQUES, VOL.~60, NO.~12, DECEMBER~2012
}{Roberg \MakeLowercase{\textit{et al.}}: High-Efficiency Diode and Transistor Rectifiers}

\maketitle

\begin{abstract}
In Internet of Things (IoT), radio frequency fingerprints (RFF) technology has been widely used for passive security authentication to identify the special emitter. However, few works took advantage of independent oscillator distortions at the receiver side, and no work has yet considered filtering receiver distortions. In this paper, we investigate the RFF identification (RFFI) involving unknown receiver distortions, where the phase noise caused by each antenna oscillator is independent. 
Three RFF schemes are proposed according to the number of receiving antennas. When the number is small, the Mutual Information Weighting Scheme (MIWS) is developed by calculating the weighted voting of RFFI result at each antenna; when the number is moderate, the Distortions Filtering Scheme (DFS) is developed by filtering out the channel noise and receiver distortions; when the number is large enough, the Group-Distortions Filtering and Weighting Scheme (GDFWS) is developed, which integrates the advantages of MIWS and DFS. Furthermore, the ability of DFS to filter out the channel noise and receiver distortions is theoretically analyzed at a specific confidence level.
Experiments are provided when both channel noise and receiver distortions exist, which verify the effectiveness and robustness of the proposed schemes.
\end{abstract}


%
\IEEEpeerreviewmaketitle

\section{Introduction}

\IEEEPARstart{T}{he} number of IoT access devices deployed in practical systems is rising quickly as a result of the expansion of IoT application scenarios including automotive industry \cite{9509294}, healthcare \cite{9311140}, smart living \cite{9389849}, etc. According to the International Data Corporation, approximately 40 billion IoT devices will be available globally by 2025. As a large number of IoT devices have plunged into human life, network security has emerged as a growing public concern worldwide \cite{9994003, yaqoob2019internet}.

Traditional cryptography-based algorithms are used in most existing wireless communication systems to achieve secure authentication of upper-layer mechanisms \cite{9483910, jan2021survey}. However, these algorithms suffer from some limitations such as computer-limited assumptions, replay attack susceptibility, high communication overhead, complexity, etc \cite {9279294}. In contrast, radio frequency fingerprints (RFF) as a very promising non-cryptographic authentication technology has recently gained a lot of research interest because of its information-theoretic security, low complexity, and high compatibility. 

The concept of RFF was first introduced in 2003 \cite{hall2003detection}. It extracts the inherent, stable, and unique fingerprints of different emitters to distinguish their physical layer properties \cite{chen2021emitter}. 
Such fingerprints exist due to the unavoidable accuracy errors and randomness in the device production process \cite{9385103}, which presents as the unintentional modulation at the emitter side, causing minor emitter distortions of the signal that are difficult to imitate.

Although this unintentional modulation is not conducive to the demodulation of the signal, it is the basis for RFF identification (RFFI) to extract radio frequency (RF) features with uniqueness, stability, and intrinsicality to complete the special emitter identification. 
The current RF features can be categorized into transient and steady-state features \cite{8970312}. For the transient features, though some scholars have confirmed their feasibility \cite{494069}, it is difficult to determine its starting point accurately due to the extremely short transient signal, which disables complete feature extraction. Therefore, a lot of researches focus on the feature extraction of the steady-state signal segment. For example, Q. Li \it{et al.} \rm{adopt} the self-phase optical function to optimize variational mode decomposition (VMD) and suppress the modal aliasing after signal decomposition \cite{9525541}. Y. Li \it{et al.} \rm{extract} features through entropy information and spectral feature method \cite{9039659}. 

In addition to RF feature extraction, RFFI contains another two steps: signal pre-processing and signal classification. The signal pre-processing, e.g., transformation \cite{9847224, 9721428}, data cleaning \cite{9980531}, etc., is used to improve the distinguishability of the subsequent extracted features among different emitters.
In terms of signal classification, some traditional classifiers, such as  K-means, support vector machine (SVM), and neural networks \cite{8970312}, are commonly adopted to classify the extracted RF features.

Although the above-mentioned works have studied various methods of each step in RFFI, few of them have taken into account receiver distortions. Such distortions unavoidably affect the accurate extraction of emitter fingerprints, and thus impact the performance of RFFI \cite{9352204}. It has been suggested to complement additional hardware to compensate for the distortions at the receiving side \cite{544464}, but it might lead to extra distortions that cannot be further explored.

Taking into account the impacts of receiver distortions and complemented hardware, B. He \it{et al.} \rm{suggest} that the performance of RFFI can be enhanced by utilizing the diversity gain of multiple received versions \cite{8757066}. 
Therefore, a configuration of multiple receiving antennas is expected to obtain similar benefits.
On the other hand, multiple-input multiple-output technology is indispensable in 5G communication. 
It is pointed out in \cite{8962162} that the large and heterogeneous antenna systems equipped with separate oscillators for each antenna, which generate oscillator distortions with independent identical distribution characteristics, are necessary for the future. 
Thus, it is obvious that the independent oscillators generate an independent phase noise to the signal received at each antenna.
Currently, no scholars have employed this independent identical distribution property in their RFFI studies.

Considering the above two facts when multiple receiving antennas are configured, we propose three schemes to enhance the robustness and recognition accuracy of RFF. 
These three schemes are respectively suited to the scenarios with different numbers of receiving antennas. 
To begin with, Mutual Information Weighting Scheme (MIWS) is proposed when the number of receiving antennas is small. The MIWS is a weighting algorithm that performs a weighted voting operation on the RFFI result at each antenna. It estimates weights based on the mutual information (MI) between the emitter and the received signal at each antenna.
Then, when the number of receiving antennas is moderate, Distortions Filtering Scheme (DFS) is proposed to filter out the channel noise and the receiver distortions by exploiting the independent identical distribution property of the received signal.
Further, the Group-Distortions Filtering and Weighting Scheme (GDFWS) is proposed to solve the performance saturation phenomenon of DFS when the number of receiving antennas is large. 
Finally, we use absolute accuracy and confidence level as the metrics of filtering ability to theoretically derive the minimum number of receiving antennas required to satisfy certain performance of DFS.
Thereby, the specific scenario in terms of the number of receiving antennas that is applicable for each scheme is derived.

The contributions of our work can be summarized as follows.
\begin{itemize}
 \item [1)]Firstly, when the number of receiving antennas is small, the MIWS scheme is proposed. It utilizes the MI between the transmitting signal and each receiving signal to measure the quality of the latter. Then, the weights of signals at each antenna are calculated accordingly to get the weighted voting of the RFFI results.
  \item [2)]Secondly, when the number of receiving antennas is moderate, the DFS scheme is proposed to deal with channel noise and receiver distortions. To our knowledge, this is the first attempt to filter out the receiver distortions in current literature.
  \item[3)]Thirdly, when the number of receiving antennas is large, the GDFWS scheme is proposed, which enjoys the advantages of both DFS and MIWS. The GDFWS uniformly divides all the antennas into groups first, then filters out channel noise and receiver distortions using DFS within each group to get robust RFFI results, and obtains the weighted voting result of all groups by MIWS. 
  \item[4)] Finally, based on the absolute accuracy and confidence level metrics, we theoretically derive the ability of DFS to filter out negative factors to determine the application scenario of DFS. The results simultaneously indicate the specific application scenarios of another two schemes.
 \end{itemize}
 
The remainder of this paper is organized as follows. Section II briefly reviews the signal distortion model and generalizes the uplink multi-antenna received RFF system model. Three RFFI schemes with multi-antenna receive are described in Section III. In Section IV, we theoretically analyze the impact of the number of receiving antennas on the performance of DFS. Section V shows the results of the experiments, followed by the conclusion in Section VI.

\section{Background and system model}
In this section, we first describe the emitter distortion, channel, and receiver distortion models, and then the uplink multi-antenna received RFF system model is established.

\subsection {Emitter distortion model}
\begin{figure*}
  \begin{center}
  \includegraphics[width=7.5in]{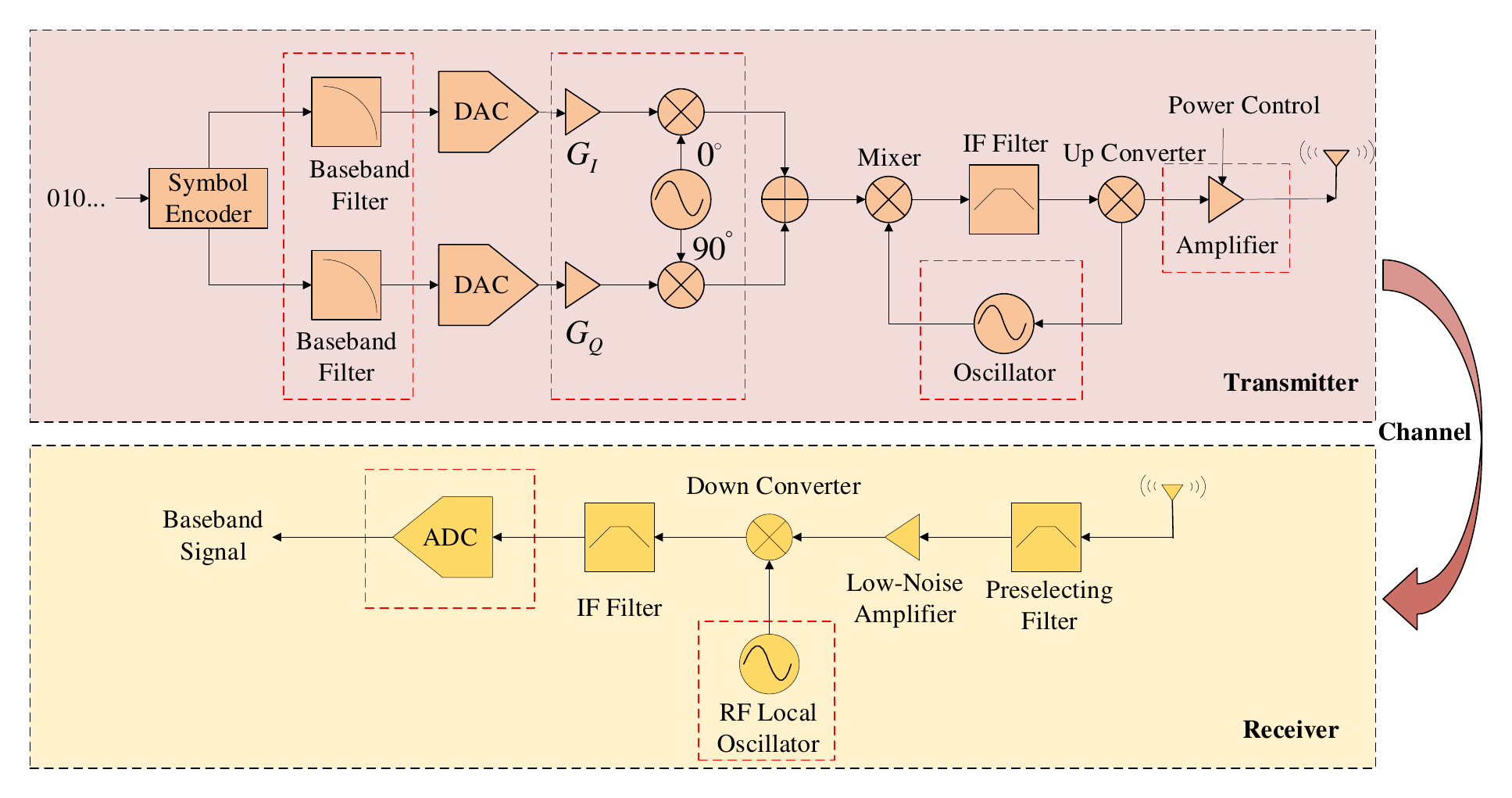}\\
  \caption{Sources of the emitter and receiver distortions during typical signal processing, where the sources of distortion modules considered in this paper are framed by red dashed lines.}\label{fingerprints}
  \end{center}
\end{figure*}
Fig.~\ref{fingerprints} depicts a typical end-to-end transceiver link and shows the source of both the emitter and receiver distortions highlighted with red dashed lines. As shown in this figure, the distortions experienced by the transmitted signal at the emitter include filter distortions, I/Q imbalance, spurious tones, and power amplifier nonlinearities. With reference to \cite{8757066} and \cite{huang2015theoretical}, the specific mathematical models for these distortions are given as follows.

\begin{itemize}
  \item [1)] Transmit shaping filter distortion:
  
We denote the $n$th transmitted symbol after constellation mapping as $S_n$, and the symbol interval is $T_s$. Subsequently, $S_n$ passes through transmit shaping filter. Considering the inevitable filter distortions due to the limited precision during the manufacturing process, the actual  transmit shaping filter is written as
\begin{equation}\label{eq1}
\widetilde{g}_t(t)=g_t(t)\otimes\upsilon(t), 
\end{equation}
where $\otimes$ stands for the convolution, $g_t(t)$ is assumed to be the ideal transmit shaping filter, and $\upsilon(t)$ denotes the filter distortion. The Fourier transform form of $\upsilon(t)$ is as follows:
\begin{equation}\label{eq2}
\Upsilon(f)=A_\Upsilon(f)e^{j\Phi_\Upsilon(f)}, 
\end{equation}
where $A_\Upsilon(f)$ and $\Phi_\Upsilon(f)$ denote the amplitude distortion and phase distortion of the filter, respectively. Current literature generally uses the second-order Fourier series model to characterize these two distortions \cite{8757066}, such that
\begin{equation}
A_\Upsilon(f)=\rho_0+\rho_1\rm{cos}(\frac{2\pi\it{f}}{\it{T_A}}), 
\end{equation}
\begin{equation}
\Phi_\Upsilon(f)=2\pi q_0f+q_1\rm{sin}(\frac{2\pi\it{f}}{\it{T}_{\rm{\Phi}}}), 
\end{equation}
where $\rho_i$, $q_i$, $i=0,1$, $T_A$, $T_\Phi$ are the parameters of the Fourier series. With the filter in (\ref{eq1}), the transmission symbols after filter shaping can be written as 
\begin{equation}\label{eq5}
s(t)=\sum_{n=-\infty}^\infty\widetilde{g}_t(t-nT_s)S_n. 
\end{equation}
  \item[2)] I/Q distortion:
  
We denote the in-phase and quadrature components of $s(t)$ in (\ref{eq5}) by $s_I(t)$ and $s_Q(t)$, respectively. The imbalance between $s_I(t)$ and $s_Q(t)$ caused by the modulator is called I/Q distortion, which is mainly manifested as a gain mismatch and quadrature error, then the signal in (\ref{eq5}) changes into
\begin{equation}\label{eq6}
\begin{aligned}
x(t)&=G_Is_I(t)e^{j(2\pi ft+\zeta/2)}+G_Qs_Q(t)e^{j(2\pi ft-\zeta/2)}\\
&=\alpha s(t)e^{j2\pi ft}+\beta s^*(t)e^{j2\pi ft}, 
\end{aligned}
\end{equation}
where $G_I$ and $G_Q$ represent the gain mismatches of these two components, $\zeta$ denotes the quadrature error, and $(\cdot)^*$ stands for conjugate operation. To facilitate the following discussion, we define
\begin{equation}
\begin{cases}
\alpha=\frac{1}{2} (G+1)\rm{cos}(\frac{\zeta}{2})+\frac{\it{j}}{2}(\it{G}-\rm{1})\rm{sin}(\frac{\zeta}{2})\\
\beta=\frac{1}{2} (G-1)\rm{cos}(\frac{\zeta}{2})+\frac{\it{j}}{2}(\it{G}+\rm{1})\rm{sin}(\frac{\zeta}{2}) .\\
G=G_I/G_Q \\
\end{cases}
\end{equation}
\item[3)] Spurious tone:

Affected by oscillators and other active devices, DC offset commonly exists in the signal. The presence of DC offset will result in harmonic components, which we refer to as spurious tone.
Considering the impact of spurious tone, the signal in (\ref{eq6}) becomes
\begin{equation}\label{eq8}
x^{(1)}(t)=x(t)+\sum_{i=1}^Cc_ie^{j2\pi (f+f_{\zeta,i})t}, 
\end{equation} 
where $C$ is the number of harmonic components, $c_i$ and $f_{\zeta,i}$ are the amplitude and frequency offset of the $i$th harmonic component, respectively. It is worth noting that when $f_{\zeta, i}=0$ and $c_i\neq 0$, the $i$th harmonic component is also known as carrier leakage.
\item[4)] Power amplifier nonlinearities:

The nonlinearity of the power amplifier causes distortions in both the amplitude and phase of the signal. We express these nonlinear distortions by the Taylor series \cite{9352204}. Considering a Taylor series of order $B$, the distorted signal in (\ref{eq8}) further becomes
\begin{equation}\label{eq9}
x^{(2)}(t)=\sum_{i=0}^Bb_i(x^{(1)}(t))^{2i+1}, 
\end{equation} 
where $b_i$ is the $i$th coefficient of the Taylor polynomial.
\end{itemize}

\subsection {Channel and receiver distortion models}
In this subsection, we consider the case of single antenna reception for simplicity. This discussion can be extended to multi-antenna reception easily, which will be described in detail in the next subsection. The channel attenuation is denoted by $h(t)$ and the additive white Gaussian noise (AWGN) is defined as $w(t)$, then the signal received by the antenna is
 \begin{equation}\label{eq10}
 z(t)=h(t)x^{(2)}(t)+w(t).
 \end{equation} 
 
In the aspect of receiver distortions, we concentrate on the phase noise caused by the oscillator, as well as the sampling jitter and quantization error caused by the ADC, which have a greater impact on the received signal compared with other hardware modules \cite{huang2015theoretical}. These receiver distortions are modeled as follows.
\begin{itemize}
  \item[1)] Phase noise:
  
Assume a phase-locked loop (PLL) is used at the receiver side for phase synchronization. As a typical signal frequency tracker, PLL has the advantages of high output stability and continuously adjustable phase, etc. However, it inevitably produces phase noise, under the impact of which the output signal of PLL is given as:
  \begin{equation}\label{eq11}
 y^{(1)}(t)=h(t)x^{(2)}(t)e^{-j(2\pi f^{\prime} t+\theta(t))}+\hat{w}(t)e^{-j(2\pi f^{\prime} t)}, 
 \end{equation} 
 where
   \begin{equation}\label{eqw}
\hat{w}(t)=w(t)e^{-j\theta(t)}, 
 \end{equation} 
$f^{\prime}$ is the local oscillator frequency and $\theta(t)$ denotes its phase noise. Similar to most studies, e.g.,  \cite{8757066, huang2015theoretical}, we model the phase noise as a Wiener process as follows:
 \begin{equation}\label{eq12}
 \theta(t)=\frac{1}{2\pi \chi}\frac{d\theta(t)}{dt}+c(t), 
 \end{equation} 
 where $\chi$ is the 3 dB bandwidth of the phase noise power spectrum and $c(t)$ is the noise obeying the standard Gaussian distribution.
 
  \item[2)] Sampling jitter:
  
Sampling jitter means the deviation of the sampling point from the optimal position when the signal is downsampled by the ADC. In the presence of sampling jitter, the signal in (\ref{eq11}) changes into
   \begin{equation}\label{eq13}
y^{(2)}(n)=y^{(1)}(nT+\delta(n)T), 
 \end{equation} 
 where $n$ denotes the $n$th sampling point, $T$ is the sampling period and $T\ll T_s$, and $\delta(n)$ denotes the relative sampling jitter, which is a random process, and $|\delta(n)|\ll 1$.
 
  \item[3)] Quantization error:
  
  The signal is quantized after sampling, and the quantization error is usually modeled as additive noise in the case of uniform quantization. The quantized version of (\ref{eq13}) is written as: 
  \begin{equation}\label{eq14}
y(n)=y^{(2)}(n)+\bigtriangleup(n), 
 \end{equation} 
 where $\bigtriangleup(n)$ denotes the quantization error at the $n$th sampling point. If quantization accuracy is  $\epsilon$ and the dynamic range is $[-V,V]$, then $\bigtriangleup (n)$ obeys a uniform distribution within the interval $[-2^{-\epsilon}V,2^{-\epsilon}V]$ with a variance of $2^{-2\epsilon}V^2/3$.
 
\end{itemize}

To summarize, based on (\ref{eq9}) to (\ref{eq14}), when the effects of the emitter distortions, channel, and receiver distortions are considered, the down-converted signal at the receiver is expressed as (\ref{eq16}) written at the top of next page, where the second equation is the simplified form of (\ref{eq16}) since sampling jitter does not affect the distribution of $\hat{w}(nT)$.
\begin{figure*}
\begin{equation}\label{eq16}
y(n)=h(nT+\delta(n)T)x^{(2)}(nT+\delta(n)T)e^{-j(\theta(nT+\delta(n)T)+2\pi f^{\prime} T\delta(n))}+\bigtriangleup(n)+\hat{w}(nT+\delta(n)T)e^{-j(2\pi f^{\prime} T\delta(n))}.
\end{equation} 
\end{figure*}
\begin{figure*}
\begin{equation}\label{eq16_1}
y(n)=h(nT+\delta(n)T)x^{(2)}(nT+\delta(n)T)e^{-j(\theta(nT)+2\pi f^{\prime} T\delta(n))}+\bigtriangleup(n)+\hat{w}(nT).
\end{equation} 
 \hrulefill
\end{figure*}
%

\begin{table*}
\caption{\label{tabI} Three Multi-antenna Received RFFI Schemes.}
\begin{center}
\setlength{\tabcolsep}{1.5mm}{
\begin{tabular}{|c|c|c|c|} \hline
Scheme&MIWS&DFS&GDFWS\\ \hline
Basic ideas&MI weighting&distortions filtering&intra-group distortions filtering, inter-group weighting\\ \hline
Underline principle&diversity gains&statistical characteristics&statistical characteristics and diversity gains\\ \hline
Application scenario&small numbers of receiving antenna&medium numbers of receving antenna&large numbers of receving antenna\\ \hline
\end{tabular}}
\end{center}
\end{table*}

\subsection {RFF system model with multiple receiving antennas}
Based on the single-antenna reception model in the previous subsection, this subsection extends it to a multi-antenna reception scenario. Assuming an uplink multi-antenna received RFF system model consists of $M$ single-antenna IoT devices and a $N$-antenna receiver, each antenna of which is equipped with an independent oscillator. 

Suppose that multiple emitters adopt orthogonal access technology to communicate with the receiver, thus this paper does not consider the interference among multiple emitters. 
According to the previous subsections, the down-converted signal at the $i$th antenna is established as:
 \begin{equation}\label{eq17}
 \begin{aligned}
y_i(k,n)=h_i(k,n)e^{-j\theta_i(k,n)}\hat{x}_m(k,n)\\
+\bigtriangleup_i(k,n)+\hat{w}_i(k,n)
\end{aligned},
 \end{equation} 
where
\begin{equation}\label{eq18}
h_i(k,n)=h_i(k,nT+\delta(k,n)T),
\end{equation} 
\begin{equation}\label{eq19}
\hat{x}_m(k,n)=e^{-j2\pi f^{\prime} T\delta(k,n)}x_m^{(2)}(k,nT+\delta(k,n)T),
\end{equation} 
 $(k,n)$ represents the $n$th sample in the $k$th frame. $h_i(k,n)$ is the channel fading coefficient between the emitter and the $i$th antenna of the receiver, $\theta_i(k,n)$ represents the phase noise of the  $i$th antenna at the receiver side, $\delta(k,n)$ denotes the sampling jitter, $\bigtriangleup_i(k,n)$ indicates the quantization error at the $i$th antenna, $\hat{w}_i(k,n)$ is the AWGN at the $i$th antenna, and $x_m^{(2)}(k,nT+\delta(n)T)$ in the form of (\ref{eq9}) for the $m$th emitter.

As mentioned in the previous subsection, the antenna oscillator inevitably generates phase noise. Fortunately, it is reasonable to assume the phase noise remains constant within a single frame while varies frame-by-frame in this paper \cite{9353800, 9229093}. 
Meanwhile, we assume a slow fading channel, so $h_i(k,n)$ remains constant within a signal frame. 
Based on these two assumptions, we define $h_i(k)\triangleq h_i(k,n)$ and $\theta_i(k)\triangleq \theta_i(k, n)$ for any $n$, and
the model in (\ref{eq17}) is further simplified as
 \begin{equation}\label{eqsys}
 \begin{aligned}
y_i(k,n)=h_i(k)e^{-j\theta_i(k)}\hat{x}_m(k,n)+\bigtriangleup_i(k,n)+\hat{w}_i(k,n)
\end{aligned}.
 \end{equation}

\section{RFFI schemes with multi-antenna receiver}
\begin{figure*}
  \begin{center}
  \includegraphics[width=7.25in]{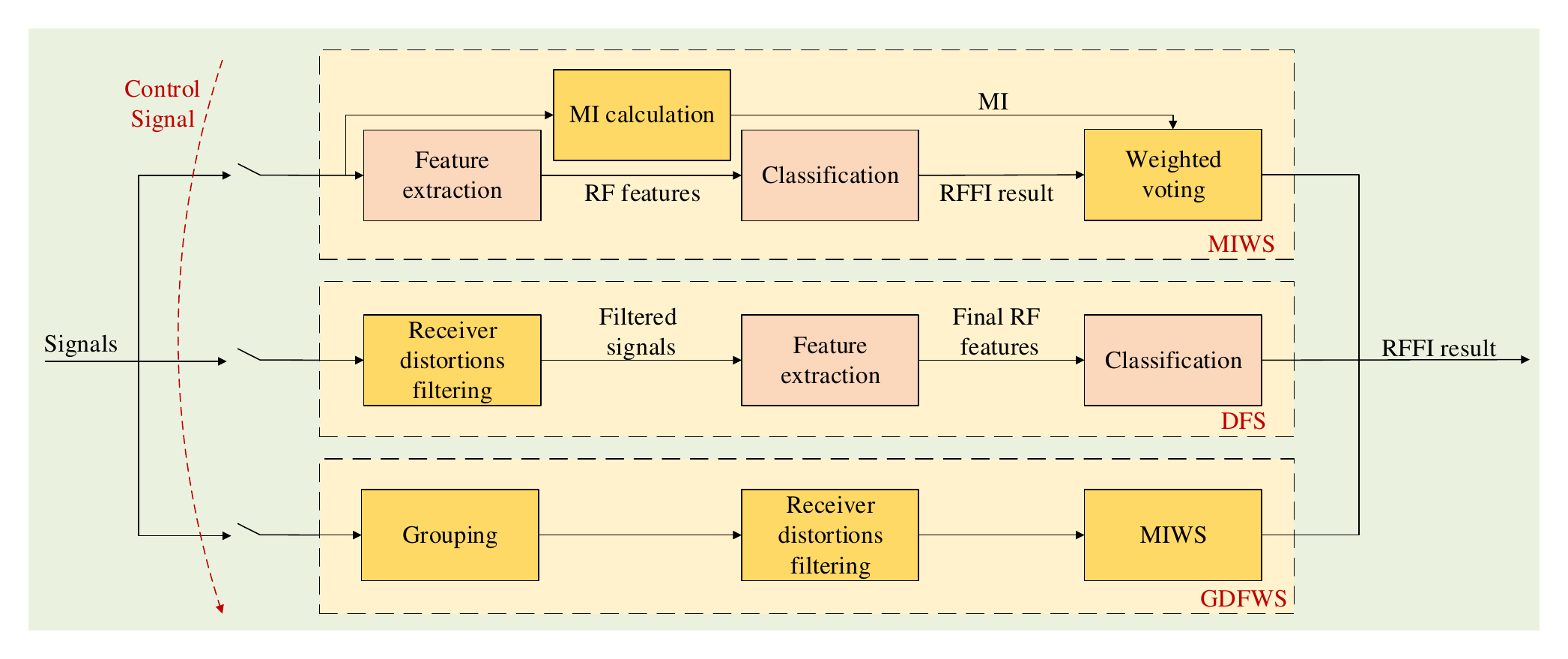}\\
  \caption{The framework of RFFI with the proposed schemes, where only one scheme can be activated at a time according to the control signal, and the modules highlighted in yellow are proposed in this paper.}\label{schemes}
  \end{center}
\end{figure*}
In this section, we propose three schemes to realize RFFI as summarized in Table I. These three schemes are applicable to scenarios with different numbers of receiving antennas. As shown in Table I, MIWS is first proposed in case the number of receiving antennas is small. Then, when the number of receiving antennas is sufficient to derive the statistical characteristics of the received signals, we propose DFS to filter out channel noise and receiver distortions. Finally, to address the issue of performance saturation that DFS encounters when the number of antennas is too large, GDFWS is proposed.
Fig.~\ref{schemes} depicts the framework of RFFI with these three schemes, where only one scheme is activated by the control signal according to the number of configured antennas, and the modules that we propose are highlighted in yellow.

\subsection {Mutual information weighting scheme}
The MI between the emitted signal and the received signal reflects their information similarity. In this paper, the larger the MI is, the less the emitter signal is affected by channel noise and receiver distortions. Based on this fact, we first estimate the MI between the emitter signal and the received signal at each antenna. Then the weight of the received signal at each antenna is set to be proportional to its corresponding MI. Finally, the RFFI result is obtained as the weighted voting of all classification results predicted from received signals.

In this subsection, we calculate the MI between the emitter signal and the received signal by taking one receiving antenna as an illustration, which can be extended to other antennas easily. For simplicity, the subscript $i$ of the $i$th antenna in the subsequent discussion is ignored.
Based on the previous analysis, we know that the emitter signal $x^{(2)}(t)$ undergoes channel fading $h(t)$, AWGN $w(t)$, and receiver distortions before completing the classification.
To facilitate subsequent analysis, we define
 \begin{equation}\label{eq20}
g(t)=h(t)x^{(2)}(t).
 \end{equation}

Firstly, with the definition in (\ref{eq20}), we transform the signal in (\ref{eq10}) into the frequency domain and get
 \begin{equation}
\bf{Z}(\it{f})=\bf{G}(\it{f})+\mathbf{W}(\it{f}),
 \end{equation}
where $\bf{Z}(\it{f})$, $\mathbf{G}(\it{f})$, and $\mathbf{W}(\it{f})$ are the expressions in frequency domain of $z(t)$, $g(t)$, and $w(t)$, respectively.
According to \cite{bell1988information}, the MI between $\mathbf{G}(\it{f})$ and $\bf{Z}(\it{f})$ is calculated as
\begin{equation}\label{eq21}
\begin{aligned}
\mathcal{I}(\bf{Z}\it{(f)};\bf{G}\it{(f)})&=\mathcal{H}(\bf{Z}\it{(f)})-\mathcal{H}(\bf{Z}\it{(f)}|\bf{G}\it{(f)})\\
&=\mathcal{H}(\bf{Z}\it{(f)})-\mathcal{H}(\bf{W}\it{(f)})\\
&=\frac{1}{2}\rm{ln}(2\pi\it{(\sigma_g^{\rm{2}}+\sigma_w^{\rm{2}})})-\rm{\frac{1}{2}ln(2\pi\it{\sigma_w^{\rm{2}})}}\\
&=\frac{1}{2}\rm{ln}(1+\it{\frac{\sigma_g^{\rm{2}}}{\sigma_w^{\rm{2}}}})
\end{aligned}, 
\end{equation}
 where $\sigma_g^2$ and $\sigma_w^2$ are the variance of $g(t)$ and $w(t)$, respectively. The function $\mathcal{H}(\cdot) $ implements entropy calculation. 

Consider a special case of (\ref{eq21}), i.e., no receiver distortion $d(t)$ exists. In this case, $\sigma_y^2=\sigma_z^2$,
where $\sigma_y^2$ represents the variance of $y(t)$. Then the special case of above equation is
  \begin{equation}\label{eq22}
  \begin{aligned}
\mathcal{I}_s(\bf{Z\it{(f)}};\bf{G}\it{(f)})&=\mathcal{I}_s(\bf{Y}\it{(f)};\bf{G}\it{(f)})\\
&=\frac{1}{2}\rm{ln}(2\pi\it{\sigma_y^{\rm{2}}})-\rm{\frac{1}{2}ln(2\pi\it{\sigma_w^{\rm{2}}})}\\
&=\frac{1}{2}\rm{ln(\frac{\it{\sigma_y^{\rm{2}}}}{\it{\sigma_w^{\rm{2}}}}})
 \end{aligned}.
 \end{equation}

Note that (\ref{eq22}) is only an ideal case, which does not exist in practical systems. To measure the quality of the received signal, we calculate the difference between (\ref{eq21}) and (\ref{eq22}) as follows,
   \begin{equation}\label{eq23}
  \begin{aligned}
\bigtriangleup I&=\mathcal{I}_s(\bf{Z}\it{(f)};\bf{G}\it{(f)})-\mathcal{I}(\bf{Z}\it{(f)};\bf{G}\it{(f)})\\
&=\frac{1}{2}\rm{ln(\frac{\it{\sigma_y^{\rm{2}}}}{\it{\sigma_w^{\rm{2}}+\sigma_g^{\rm{2}}}}})\\
&=\frac{1}{2}\rm{ln(\frac{\it{\sigma_y^{\rm{2}}}}{\it{\sigma_w^{\rm{2}}+\sigma_{x^{\rm{(2)}}}^{\rm{2}}\sigma_h^{\rm{2}}}}})\\
 \end{aligned},
 \end{equation} 
where $\sigma_h^2$ and $\sigma_{x^{\rm{(2)}}}^{\rm{2}}$ indicate the variance of $h(t)$ and $x^{(2)}(t)$ in (\ref{eq10}), respectively. In practical applications, $\sigma_w^2$ and $\sigma_h^2$ can be derived by some Signal to Noise Ratio (SNR) estimation techniques \cite{9075273, 901810, 8461787}, and channel coefficient estimation techniques\cite{7996889, 8665885}. Additionally, $\sigma_{x^{(2)}}^2$ can also be estimated based on the received pilots.

Obviously, a smaller $\bigtriangleup I$ indicates fewer distortions at the receiver. Based on this fact, we define the weight of $y_i(t)$ as,
   \begin{equation}\label{eqweight}
\omega_i=\frac{1/\bigtriangleup I_{\it{i}}}{\sum_{j=1}^N1/\bigtriangleup I_{\it{j}}},
 \end{equation}  
where $\bigtriangleup I_i$ represents the MI difference at the $i$th receiving antenna in the form of (\ref{eq23}).
 
 We use $s\it{_i}$ to denote the classification result of $y_i(t)$. Then the weighted voting of all results is implemented based on their corresponding weights in (\ref{eqweight}), and the final RFFI result can be obtained.
\subsection {Distortions filtering scheme}
Though the MIWS described in the previous subsection realizes the reduction of the negative impact of channel noise and receiver distortions through diversity gain, a more direct way is to eliminate these negative effects.
Thus, this subsection proposes DFS to filter out the channel noise and receiver distortions when the number of receiving antennas is sufficiently large.

If not specifically mentioned, the following derivations are given for single-frame signal, thus we omit the label $k$.
By defining $\phi_i=h_ie^{-j\theta_{i}}$, the overall system model in (\ref{eqsys}) is rewritten as
\begin{equation}\label{eq26}
y_i(n)=\phi_i\hat{x}_m(n)+\bigtriangleup_i(n)+\hat{w}_i(n).
 \end{equation} 
To improve the quality of  $y_i(n)$, DFS attempts to filter out the adverse factors, i.e., $\phi_i$, $\bigtriangleup_i(n)$, and $\hat{w}_i(n)$.
 
First, by considering all the sampling points of all antennas, (\ref{eq26}) is converted into the matrix form
 \begin{equation}
\begin{aligned}
\bf{Y}&=\bf{\Phi}\bf{\hat{x}}^{\it{T}}+\Delta+\bf{\hat{W}}\\
&=\bf{\Xi}+\Delta+\bf{\hat{W}}
\end{aligned},
 \end{equation} 
where
 \begin{equation}
 \bf{\Phi}=\
 \begin{pmatrix}
\phi_1&\phi_2&\cdots&\phi_N
\end{pmatrix}
^{\it{T}}\in\mathbb{R}^{\it{N}\times\rm{1}}, 
 \end{equation}  
 \begin{equation}
 \bf{\hat{x}}=\
 \begin{pmatrix}
\hat{x}_m(1)&\hat{x}_m(2)\cdots&\hat{x}_m(L)
\end{pmatrix}
^{\it{T}}\in\mathbb{R}^{\it{N}\times\rm{1}}, 
 \end{equation} 
 \begin{equation}\label{eqxphi}
 \bf{\Xi}=\bf{\Phi}\bf{\hat{x}}^{\it{T}}\in\mathbb{R}^{\it{N}\times N},
 \end{equation}
 $\Delta\in\mathbb{R}^{\it{N}\times L}$ with the $n$th element of the $i$th row being $\bigtriangleup_i(n)$, $\bf{\hat{W}}\in\mathbb{R}^{\it{N}\times L}$ with the $n$th element of the $i$th row being $\hat{w}_i(n)$, and $L$ is the number of sampling points in a frame. 
 
 The basic idea of DFS is to recover the matrix $\bf{\Xi}$ from $\bf{Y}$, and then recover $\bf{\hat{x}}$ based on its relationship with $\bf{\Xi}$ given in (\ref{eqxphi}). In doing so, the impacts of $\bf{\Phi}$, $\Delta$, and $\bf{\hat{W}}$ are expected to be eliminated.
 
To  better illustrate the statistical properties of the received signals, the matrix $\bf{Y}$ is rewritten as (\ref{eq33}) shown at the top of the next page,
\begin{figure*}
  \begin{center}
    \begin{equation}\label{eq33}
     \begin{aligned}
     \bf{Y}&=
     \begin{pmatrix}
    \phi_1\cdot(\hat{x}_m(1)+v_{11})&\phi_1\cdot(\hat{x}_m(2)+v_{12})&\cdots&\phi_1\cdot(\hat{x}_m(L)+v_{1L})\\
    \phi_2\cdot(\hat{x}_m(1)+v_{21})&\phi_2\cdot(\hat{x}_m(2)+v_{22})&\cdots&\phi_2\cdot(\hat{x}_m(L)+v_{2L})\\
    \vdots&\vdots&\ddots&\vdots\\
    \phi_N\cdot(\hat{x}_m(1)+v_{N1})&\phi_N\cdot(\hat{x}_m(2)+v_{N2})&\cdots&\phi_N\cdot(\hat{x}_m(L)+v_{NL})\\
    \end{pmatrix}, 
    \end{aligned}
 \end{equation} 
 \end{center}
\end{figure*}
where
 \begin{equation}\label{mean0}
 v_{ij}=(\hat{w}_i(j)+\bf{\bigtriangleup}_{\it{i}}\it{(j)})/\phi_{\it{i}}.
 \end{equation}
It should be noted that $\bf{\bigtriangleup}_{\it{i}}\it{(j)}$ and $\hat{w}_i(j)$ follow the uniform and Gaussian distribution, respectively. Both of them have a mean of 0. Hence, the mean of $v_{ij}$ is 0. Based on this property, we calculate the mean of each row of the matrix in (\ref{eq33}) to obtain
\begin{equation}\label{eq35}
\mathcal{E}(\bf{Y}_{\it{i},.})=\phi_{\it{i}}\mathcal{E}(\bf{\hat{x}}^{\it{T}}+\bf{v}_{\it{i},.})=\phi_{\it{i}}\mathcal{E}(\bf{\hat{x}}), 
\end{equation} 
where $\bf{Y}_{\it{i},.}$ denotes the $i$th row of the matrix $\bf{Y}$ in (\ref{eq33}), 
\begin{equation}
\bf{v}_{\it{i},.}=
\begin{pmatrix}
v_{i1}&v_{i2}&\cdots&v_{iL}
\end{pmatrix},
\end{equation}
and $\mathcal{E}(\bf{u})$ calculates the mean of the vector $\bf{u}$.
 Based on (\ref{eq35}), it is easy to obtain
  \begin{equation}\label{eq37}
\frac{\mathcal{E}(\bf{Y}_{\it{i},.})}{\mathcal{E}(\bf{Y}_{\it{j},.})}=\frac{\phi_i}{\phi_j}. 
 \end{equation}
 
 Next, we reconstruct the $l$th row of $\bf{\Xi}$ from $\bf{Y}$, the processes of which apply to the other rows of $\bf{\Xi}$ as well.
By multiplying the $j$th row of (\ref{eq33}) by ${\phi_l}/{\phi_j}$ with $j=1,2,...,N$, which has been obtained by (\ref{eq35}) and (\ref{eq37}), the matrix in (\ref{eq33}) becomes (\ref{eqerr}).
\begin{figure*}
  \begin{equation}\label{eqerr}
  \widetilde{\bf{Y}}^{(l)}=\phi_l\cdot
 \begin{pmatrix}
\hat{x}_m(1)+v_{11}&\hat{x}_m(2)+v_{12}&\cdots&\hat{x}_m(L)+v_{1L}\\
\hat{x}_m(1)+v_{21}&\hat{x}_m(2)+v_{22}&\cdots&\hat{x}_m(L)+v_{2L}\\
\vdots&\vdots&\ddots&\vdots\\
\hat{x}_m(1)+v_{N1}&\hat{x}_m(2)+v_{N2}&\cdots&\hat{x}_m(L)+v_{NL}\\
\end{pmatrix}. 
 \end{equation}
 \hrulefill
 \end{figure*}
 With the derived $\widetilde{\bf{Y}}^{(l)}$ in (\ref{eqerr}), by calculating its column mean, we get
  \begin{equation}\label{eq38}
  \mathcal{E}_c(\widetilde{\bf{Y}}^{(l)})=\phi_l
  \begin{pmatrix}
\hat{x}_m(1)&\hat{x}_m(2)&\cdots&\hat{x}_m(L)
\end{pmatrix},
  \end{equation}   
where $\mathcal{E}_c(\bf{U})$ calculates the column mean of the matrix $\bf{U}$. It is worth noting that $\mathcal{E}_c(\widetilde{\bf{Y}}^{(l)})$ is exactly the $l$th row of $\bf{\Xi}$, which is denoted as $\bf{\Xi}_{\it{l},\cdot}$.
Following the above procedures, we are able to recover all the rows of $\bf{\Xi}$ by varying the value of $l$ from 1 to $N$ in order.
 
By observing $\bf{\Xi}$, we find that directly separating it into $\bf{\Phi}$ and $\bf{\hat{x}}$ without any prior knowledge of $\bf{\hat{x}}$ is impossible. Fortunately, by defining $\rm{\hat{\it{x}}_{\it{m}}(1)}$ as the first symbol of this frame, we have
\begin{equation}\label{eq39}
\bf{\Xi}=\bf{\Phi}\hat{x}^{\it{T}}=\rm{\hat{\it{x}}_{\it{m}}(1)}\bf{\Phi}\widetilde{\bf{x}}^{\it{T}},
\end{equation}
where
\begin{equation}
 \begin{aligned}
\widetilde{\bf{x}}&=\bf{\hat{x}}/\rm{\hat{\it{x}}_{\it{m}}(1)}\\
&=
 \begin{pmatrix}
1&\hat{x}_m(2)/\rm{\hat{\it{x}}_{\it{m}}(1)}&\cdots&\hat{x}_m(L)/\rm{\hat{\it{x}}_{\it{m}}(1)}
\end{pmatrix}\\
&=\frac{\sum_{i=1}^{N}\bf{\Xi}_{\it{i,.}}/\bf{\Xi}_{\it{i},\rm{1}}}{N}
 \end{aligned}.
\end{equation}
It is obvious that $\widetilde{\bf{x}}$ is highly correlated with $\bf{\hat{x}}$. 

Since $\widetilde{\bf{x}}$ retains all the information of $\bf{\hat{x}}$, it is reasonable to use $\widetilde{\bf{x}}$ rather than $\bf{\hat{x}}$ as the signal for the subsequent RF feature extraction and classification without affecting the performance of RFFI.

We find that in (\ref{eq35}) and (\ref{eq38}) the calculation of the mean is implemented. However, in practical scenarios, it can only be approximated by averaging. To ensure the effect of filtering out channel noise and receiver distortions, $L$ and $N$ should be large enough. Generally, $L$ is sufficiently large in practical applications. Therefore, the filtering ability of DFS mainly depends on the value of $N$, and their relationship will be analyzed in detail in Section IV.

\subsection{Group-distortions filtering and weighting scheme}
The previous analysis has suggested the larger $N$ is, the smaller the difference between the averaging result and the actual mean 0. 
When $N$ is sufficiently large, this difference will be definitely very small. At this moment, even if we further increase $N$, the difference will converge and thus no significant performance enhancement will appear. We call such phenomenon of DFS as performance saturation.
To alleviate this problem when the number of receiving antennas is large, this subsection proposes GDFWS that divides all antennas into several groups to avoid the appearance of saturation phenomenon.

\begin{figure*}[htbp]
  \centerline{\includegraphics[width=1\textwidth]{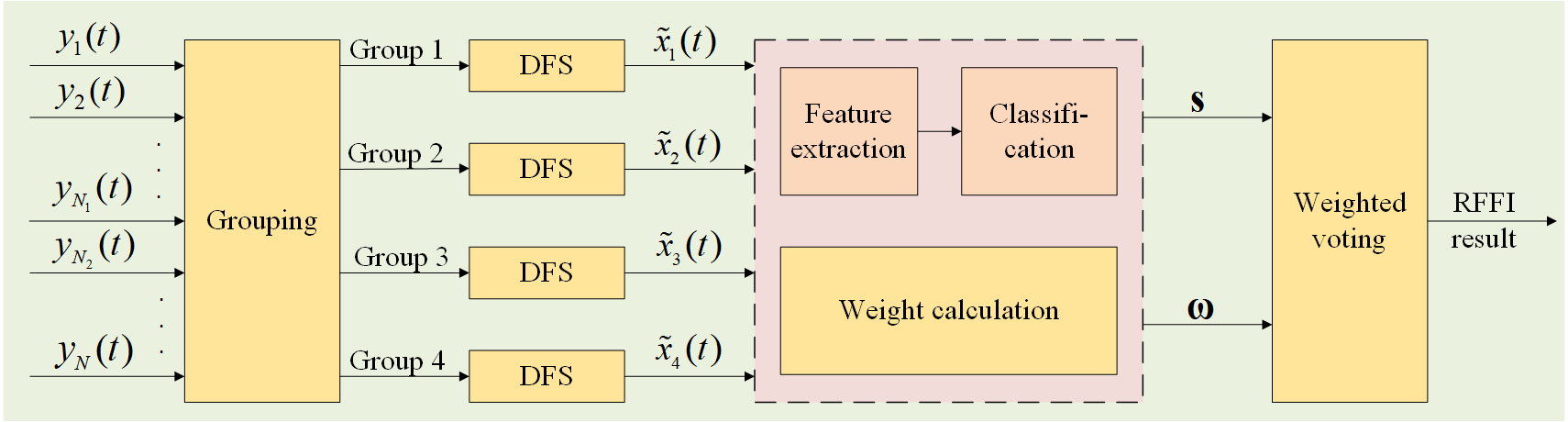}}
  \caption{\centering{GDFWS diagram when received signals are divided into two blocks, where the steps filled with yellow are proposed by us, while the pink is based on existing methods.}}\label{BDFWS}
\end{figure*}
Fig.~\ref{BDFWS} illustrates the overall structure of GDFWS, where signals received by all antennas are divided into four groups for illustration. In this figure, $N_1=N/2$, and $N_2=N/2+1$.
First, DFS is applied in each group to filter out channel noise and receiver distortions.
Then, the obtained $\widetilde{x}_i(t), i\in{1,2,3,4}$ are delivered to the feature extraction module and weight calculation module, where the former extracts RF features that are subsequently fed into the classification module to obtain classification results $\bf{s}\in\mathbb{R}^{\rm{1}\times\it{N}}$, and the later calculates their respective weights $\bf{\omega}\in\mathbb{R}^{\rm{1}\times\it{N}}$.
Finally, the classification result $\bf{s}$ and its corresponding weight $\bf{\omega}$ are sent into the weighted voting module to obtain the final RFFI result.

Obviously, GDFWS enjoys the advantages of both MIWS and DFS in terms of diversity gain and adverse factors elimination. Meanwhile, the above structure can be easily extended to the case of multiple groups, where the number of antennas in each group is smaller than the one when DFS exhibits performance saturation.

\section{Theoretical analysis of DFS and Applicable scenario discussion}
In this section, we theoretically analyze the ability of DFS to filter out channel noise and receiver distortions with varying numbers of receiving antennas. The metrics that we consider are confidence level and absolute accuracy. By revealing the relationship between $N$ and these metrics, we obtain the conclusion about which scheme is more desirable for the case with different numbers of receiving antennas.

We consider the asymptotic case, where $L\to +\infty$, and then study the approximation degree of using averaging operation instead of the actual mean, which reveals its dependence on $N$.
By averaging each column of (\ref{eqerr}), the statistical average of the $k$th column is derived as:
\begin{equation}\label{ei}
\begin{aligned}
\mathcal{E}(\widetilde{\bf{Y}}^{(l)}_{.,k})&=\phi_l\cdot\mathcal{E}
 \begin{pmatrix}
\hat{x}_m(k)+v_{1k} &\cdots&\hat{x}_m(k)+v_{Nk} 
\end{pmatrix}
^T \\
&=\phi_l(\hat{x}_m(k))+\tau_k
\end{aligned},
\end{equation}
where
 \begin{equation}\label{eq43}
 \tau_k=\mathcal{E}
  \begin{pmatrix}
\phi_lv_{1k} &\phi_lv_{2k}&\cdots&\phi_lv_{Nk} 
\end{pmatrix}.
 \end{equation}
In the above equation, $\tau_k$ represents the difference between the average result and the mean 0. Therefore, the smaller $\tau_k$ is, the better the filtering effect of DFS on adverse factors, i.e., channel noise and receiver distortions.
For simplicity, we define
\begin{equation}
u_{ik}=\phi_lv_{ik}\\,
\end{equation}
where $i\in1,2, ... , N$.
According to (\ref{mean0}), the $u_{ik}$ in the above equation can be rewritten as
\begin{equation}
\begin{aligned}
u_{ik}&=\hat{w}_i(k)+\bf{\bigtriangleup}_{\it{i}}\it{(k)}\\
&=w_i(k)e^{-j\theta_i(k)}+\bf{\bigtriangleup}_{\it{i}}\it{(k)}
\end{aligned},
\end{equation}
where the definition of each variable remains the same as in Section II.B. To further simplify, in the following analysis, we ignore $k$ in the above equation, which reduces to
\begin{equation}
u_{i}=w_ie^{-j\theta_i}+\bf{\bigtriangleup}_{\it{i}}.
\end{equation}

To obtain the distribution of $\tau$, i.e., $\tau_k$ in (\ref{eq43}), we first analyze the distribution of $u_i$. Referring to Section II.B, we know that $\bf{\bigtriangleup}_{\it{i}}$ is uniformly distributed between $[-2^{-\epsilon}V,2^{-\epsilon}V]$. 
Suppose the number of quantization bits is 16, i.e., $\epsilon=16$, and $V=1$, we have 
\begin{equation}
\bigtriangleup_{\it{i}}\sim U[-2^{-16},2^{-16}].
\end{equation}
The following discussions can be easily extended to the other cases of $V$ and $\epsilon$.
On the other hand, $w_i$ obeys a Gaussian distribution with mean 0 and variance $\sigma_w^2$, and $\theta_i$ obeys a standard Gaussian distribution, so we easily obtain
\begin{equation}
w_ie^{-j\theta_i}\sim \mathcal{N}(0,\sigma_w^2).
\end{equation}
According to (\ref{eq10}), we have 
\begin{equation}
 \sigma_w^2=\sigma_{x^{(2)}}^2\sigma_h^2/10^{\rm{SNR/10}},
\end{equation}
where $\sigma_{x^{(2)}}^2$ and $\sigma_h^2$ are the same as defined in Section III.A. For better illustration, we assume that $\sigma_{x^{(2)}}^2=1$ and $\sigma_h^2=1$, thus the above equation is further simplified as
\begin{equation}
 \sigma_w^2=10^{-\rm{SNR/10}}.
\end{equation}
When $0 \rm{dB}\le SNR\le 30 \rm{dB}$, it is clear that
\begin{equation}
2^{-16}\ll 10^{-3}\le \sigma_w^2 \le1,
\end{equation}
thus
\begin{equation}
u_{i}=w_ie^{-j\theta_i}+\bigtriangleup_{\it{i}}\approx w_ie^{-j\theta_i}\sim \mathcal{N}(0,\sigma_w^2).
\end{equation}

Considering that the average of $N$ possible samples selected randomly from $u_i, i\in1,2,...,N$ also follows the Gaussian distribution, so $\tau$ in (\ref{eq43}) obeys the Gaussian distribution. Furthermore, based on the sampling properties of the sample means \cite{ross2017introductory}, it is known that
\begin{equation}
\tau\sim \mathcal{N}(0,\frac{\sigma_w^2}{N}).
\end{equation}

Next, we discuss two performance metrics, i.e., confidence level $\alpha$ and absolute accuracy $\xi$. If $\tau$ is required to be less than $\xi$ with a confidence level $\alpha$, we have
\begin{equation}
P(|\tau|<\xi)=\int_{-\xi}^\xi\frac{\sqrt{N}}{\sqrt{2\pi}\sigma_w}e^{-\frac{\tau^2N}{2\sigma_w^2}}{\rm d}\it{\tau}=\alpha.
\end{equation}
Let $\widetilde{\tau}=\sqrt{N}\tau/\sqrt{2}/\sigma_w$, then the above equation is rewritten as
\begin{equation}
P(|\widetilde{\tau}|<a)=\int_{-a}^a\frac{1}{\sqrt{\pi}}e^{-\widetilde{\tau}^2}{\rm d}\it{\widetilde{\tau}}=\rm{erf}^{-1}(\it{a})=\alpha,
\end{equation}
where $a=\sqrt{N}\xi/\sqrt{2}/\sigma_w$. As a result, the mathematical relationship between the confidence level $\alpha$, absolute accuracy $\xi$, and the number of receiving antennas $N$ is
\begin{equation}\label{eq56}
\begin{aligned}
\xi^{\rm{2}}&=2[\rm{erf}^{-1}(\it{\alpha})]^{\rm{2}}\frac{\sigma_w^{\rm{2}}}{N}\\
&=2[\rm{erf}^{-1}(\it{\alpha})]^{\rm{2}}N^{\rm{-1}}\rm{10}^{-\rm{SNR}/10}
\end{aligned}.
\end{equation}

Clearly, a smaller $\xi$ indicates that DFS is more effective at filtering the adverse factors. 
To measure the advantage of DFS, we define the performance gain $p$ such that
\begin{equation}
p=\frac{\xi_1-\xi}{\xi_1}=1-\sqrt{\frac{1}{N}},
\end{equation}
where 
\begin{equation}
\xi_1=\sqrt{2}\rm{erf}^{-1}(\it{\alpha})\sigma_w^{\rm{2}}.
\end{equation}
Only when the gain $p$ is larger than a threshold $p_0$, we regard that introducing DFS can bring benefits.
That is
\begin{equation}\label{eq58}
p=1-\sqrt{\frac{1}{N}}>p_0.
\end{equation}
In this paper, we set $p_0=1/2$, and get $N>4$ based on the above equation. It means that when $N\le4$, no obvious gain can be obtained when DFS is employed.
In such a scenario, MIWS serves as an alternative.

Table II presents the relationship between $N$ and $\xi$ in (\ref{eq56}) when confidence level $\alpha=0.95$ and SNR=15dB.
\begin{table*}
\caption{\label{tabI} The Relationship Between $N$ and $\xi$ when $\alpha=0.95$ and SNR=15dB.}
\begin{center}
\setlength{\tabcolsep}{6mm}{
\begin{tabular}{|c|c|c|c|c|c|c|c|c|} \hline
$N$&4&8&16&32&64&128&256&512\\ \hline
$\xi$&0.1743&0.1232&0.0871&0.0616&0.0436&0.0308&0.0218&0.0154\\ \hline
$\bigtriangleup\xi$&$-$&0.0510&0.0361&0.0255&0.0180&0.0128&0.0090&0.0064\\ \hline
Scheme&MIWS&\multicolumn{5}{c|}{DFS}&\multicolumn{2}{c|}{GDFWS}\\ \hline
\end{tabular}}
\end{center}
\end{table*}
From this table, we note that the decreasing rate of $\xi$, i.e., $\bigtriangleup\xi$, slows down as $N$ increases. When $N>128$, the decreasing rate of $\xi$ is much slower than that when $N \le 128$. 
This coincides with the expectation in the previous subsection, i.e., the performance of DFS will be saturated when $N$ is large enough.
To avoid this saturation phenomenon, we use the GDFWS scheme in the cases with medium to high SNR to divide the antenna set into groups of no more than 128 antennas each, which avoids the saturation of DFS in each group.

\begin{table*}
\caption{\label{tabI} The Relationship Between $N$ and $\xi$ when $\alpha=0.95$ and SNR=5dB.}
\begin{center}
\setlength{\tabcolsep}{6mm}{
\begin{tabular}{|c|c|c|c|c|c|c|c|c|} \hline
$N$&4&8&...&256&512&1024&2048&4096\\ \hline
$\xi$&0.5510&0.3897&...&0.0689&0.0487&0.0344&0.0244&0.0172\\ \hline
$\bigtriangleup\xi$&$-$&0.1614&...&0.0285&0.0202&0.0143&0.0101&0.0071\\ \hline
Scheme&MIWS&\multicolumn{6}{c|}{DFS}&\multicolumn{1}{c|}{GDFWS}\\ \hline
\end{tabular}}
\end{center}
\end{table*}
Table III gives the relationship between $N$ and $\xi$ in low SNR, where $\alpha=0.95$ and SNR=5dB.
It is noted that the saturation of DFS appears when $N>2048$, which is much larger than that in Table II. This means that DFS is more likely to saturate with a smaller $N$ when SNR is higher. In practical applications, the number of receiving antennas is generally limited, i.e., it will not reach 2048, thus DFS is still preferable rather than GDFWS at low SNR.
\section{Experiment and discussion}
In this section, some experiments are provided to verify the efficiency of the proposed schemes.

\subsection{Experimental setting}
According to the produces of RFF, we describe the settings of the simulation experiments in this paper from five aspects in turn: emitters, channel, receiver, RF feature extraction methods, and classifiers, the detail of which are as follows.
\begin{itemize}
\item[1)] The settings of emitters:
\begin{table*}
\caption{\label{tab1} The Parameters Settings of Emitters.}
\begin{center}
\setlength{\tabcolsep}{6mm}{
\begin{tabular}{|c|ccccc|} \hline
\diagbox{P}{E}&$\rm{T_1}$&$\rm{T_2}$&$\rm{T_3}$&$\rm{T_4}$&$\rm{T_5}$\\ \hline
$\rho_0$&1&1&1&1&1\\ \hline
$\rho_1$&0.03&0.06&0.085&0.073&0.040\\ \hline
$T_A$&4&4&4&4&4\\ \hline
$q_0$&1&1&1&1&1\\ \hline
$q_1$&0.0302&0.0295&0.0290&0.0310&0.0313\\ \hline
$T_\Phi$&4&4&4&4&4\\ \hline
$G$&0.998&1.0056&1.0102&0.9992&0.9982\\ \hline
$\zeta$&-0.018$^\circ$&0.0175$^\circ$&0.012$^\circ$&0.003$^\circ$&0.024$^\circ$\\ \hline
$c_1$&0.0013+0.0082j&0.0015+0.0072j&0.0011+0.0068j&0.0017+0.009j&0.002+0.0065j\\ \hline
$c_2$&0.0082&0.0075&0.0070&0.0087&0.0090\\ \hline
$f_{\zeta,1}$&0&0&0&0&0\\ \hline
$f_{\zeta,2}$&0.0129&0.0132&0.0123&0.0135&0.0119\\ \hline
$b_1$&1&1&1&1&1\\ \hline
$b_3$&0.3&0.6&0.01&0.4&0.08\\ \hline
\end{tabular}}
\end{center}
\end{table*}\\
The RF signal is generated according to the emitter distortion model described in Section II. It should be noted that we assume the number of harmonic components in (\ref{eq8}) and the order of the Taylor series in (\ref{eq9}) both are 2. We consider 5 emitters with the distortion parameters provided in Table IV, where the E and P are abbreviations for Emitters and Parameters, respectively, and $\rm{T_1}$ to $\rm{T_5}$ labels for emitter 1 to emitter 5. The modulation mode of the RF signal is QPSK, with the oversampling factor $T/T_s=10$, $1/T_s=10^{6}$ MHz, and the signal center frequency is 1GHz. A frame consists of 128 symbols, wherein 32 symbols carry pilots.
\item[2)] The settings of channel:\\
AWGN channel is considered in our experiments, so the channel fading coefficients $h_i(k)$ in (\ref{eqsys}) is set to be 1. Nonetheless, the following experimental conclusions also apply to the case where the channel coefficients are random.
\item[3)] The settings of the receiver:\\
The receiver distortions are generated according to the receiver distortion model described in Section II. The distortion parameters of sampling jitter and quantization errors are set as follows: $\delta(n)=0.003$, $V=1$, $\epsilon=16$. The parameter $\chi$ in (\ref{eq12}) for phase noise varies to show its influence on RFFI accuracy. 
\item[4)] RF feature extraction methods:\\
Two classical RF feature extraction methods, i.e., least mean square (LMS)-based feature extraction \cite {9519652} and intrinsic time-scale decomposition (ITD)-based feature extraction \cite {8757066}, are used in our experiments. 
The LMS-based feature extraction method updates its filter weights recursively based on some criteria until convergence, then uses its converged weight vector as features. 
The ITD-based feature extraction method first decomposes the signal by ITD and then calculates the skewness and kurtosis of each decomposed signal as the feature vector.
\item[5)] Classifiers:\\
Currently, there have been many successful classifiers applied to RFFI. However, since classifiers are not the focus of our paper, we choose the typical multi-classification SVM for RFF classification.
\end{itemize}

In the following experiments, the number of training frames and testing frames for each emitter is 200 and 100, respectively. Each experiment result is obtained by averaging over 1000 trials. We use ORS to represent the original scheme without distortion filtering and the weighted voting operation, which is used as a benchmark for the proposed schemes. 
\subsection{Experimental results of MIWS}
\begin{figure}[htbp]
  \centerline{\includegraphics[width=0.5\textwidth]{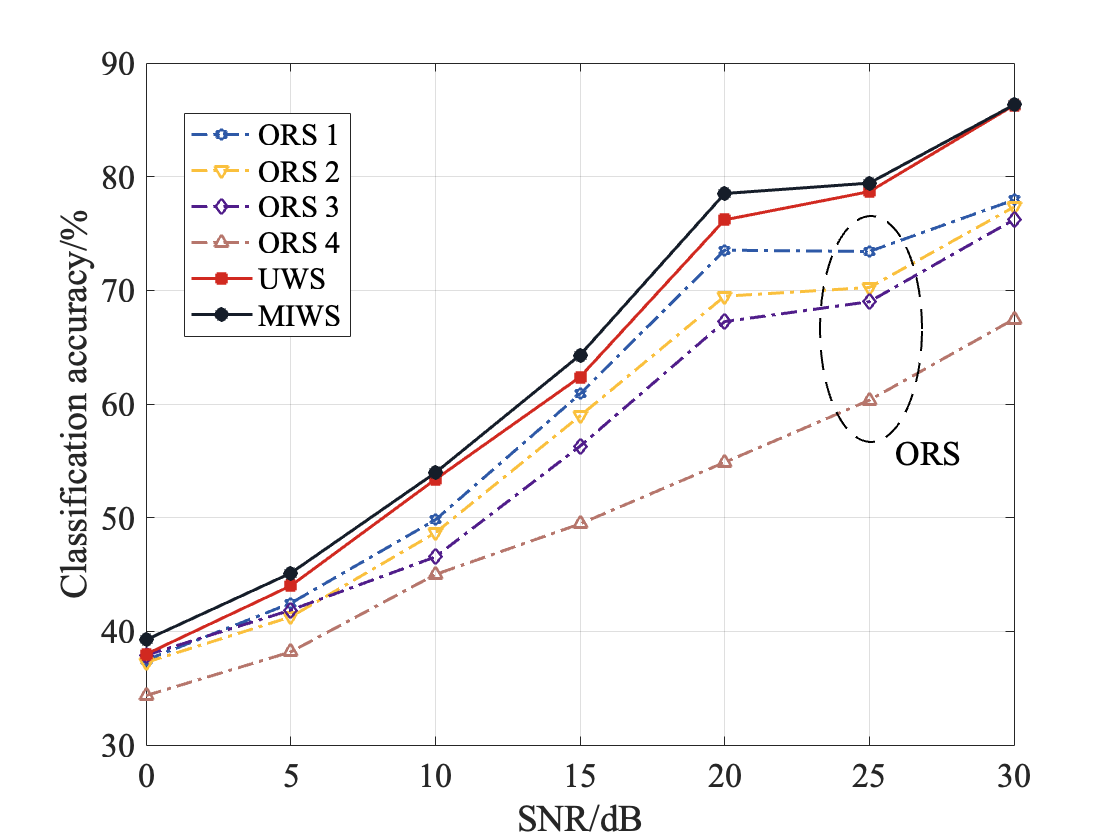}}
  \caption{\centering{Classification accuracy of MIWS based on ITD with $M=5$, $N=4$, and $\chi$ of each antenna being 0.001, 0.01, 0.1, 1 (top to bottom), respectively.}}
  \label{MIWS_3}
\end{figure}
Fig.~\ref{MIWS_3} depicts the results of RFFI accuracy with varying SNR for MIWS, where five emitters are present, the ITD-based feature extraction method is adopted, and the receiver is equipped with four antennas with $\chi$ of 0.001, 0.01, 0.1, 1, respectively. 
In this figure, UWS stands for the scheme of equally weighting the signals received by each antenna, while ORS $i$ represents the scheme that directly uses the signals at the $i$th receiving antenna without any distortion filtering or weighting operation.
It is noteworthy that the results of each ORS in this figure reveal that the larger the $\chi$, the lower the recognition accuracy, which indicates that the ITD-based feature extraction method is sensitive to the phase noise at the receiver.
Furthermore, the experimental results indicate that both MIWS and UWS outperform the ORS of each antenna, while the performance of MIWS is better than that of UWS, which highlights the benefit of setting weights according to MI.

\begin{figure}[htbp]
  \centerline{\includegraphics[width=0.5\textwidth]{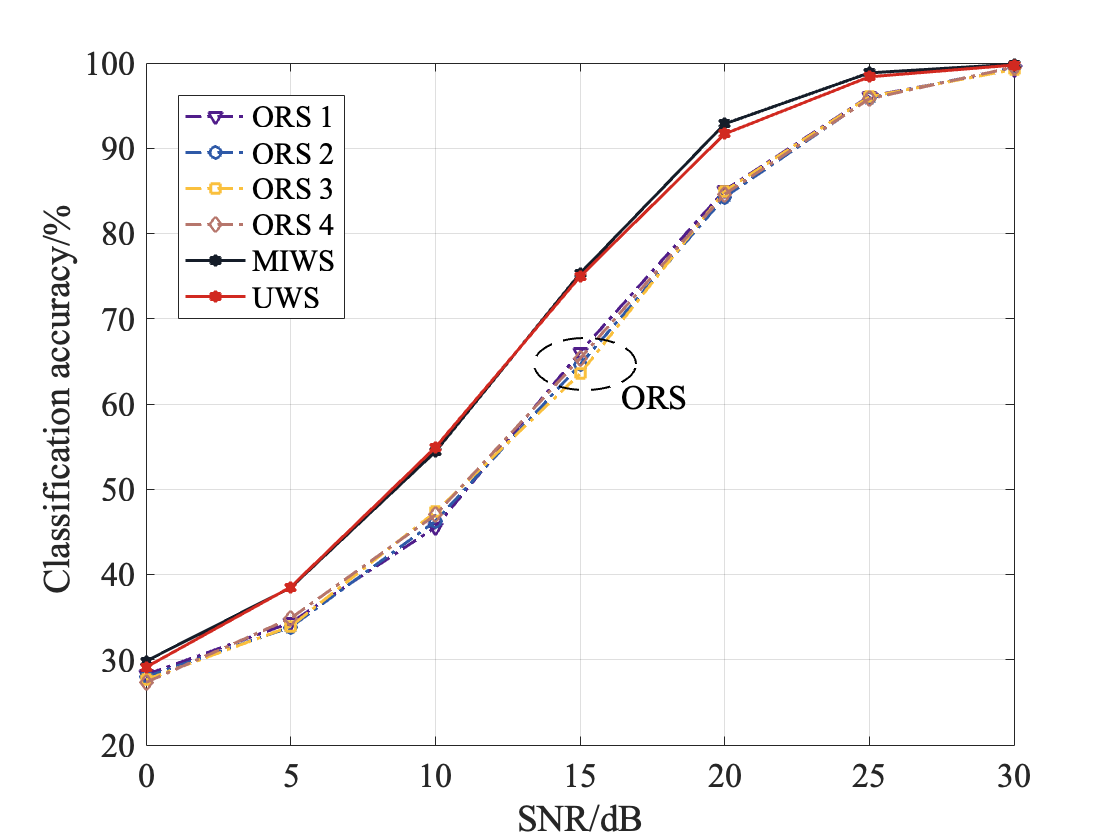}}
  \caption{\centering{Classification accuracy of MIWS based on LMS with $M=5$, $N=4$, and $\chi$ of each antenna being 0.001, 0.01, 0.1, 1, respectively.}}
  \label{MIWS_1_2}
\end{figure}
Fig.~\ref{MIWS_1_2} depicts the results of RFFI accuracy with varying SNR for MIWS, where the LMS-based feature extraction method is employed, and the other settings are the same as those in Fig.~\ref{MIWS_3}.
Unlike the ORS with significant differences in Fig.~\ref{MIWS_3}, the ORS of each antenna maintains similar RFFI accuracy when $\chi$ varies in this figure. This observation suggests that the ITD-based feature extraction method is sensitive to phase noise at the receiver, whereas the LMS-based feature extraction method is robust to it. Therefore, there is no significant difference in the performance of MIWS and UWS when the LMS-based feature extraction method is applied.
We also note that both MIWS and UWS enjoy a 10\% accuracy gain when compared with ORS,  thereby confirming the benefit of weighting when multiple received versions are configurated.
\subsection{Experimental results of DFS}
\begin{figure}[htbp]
  \centerline{\includegraphics[width=0.5\textwidth]{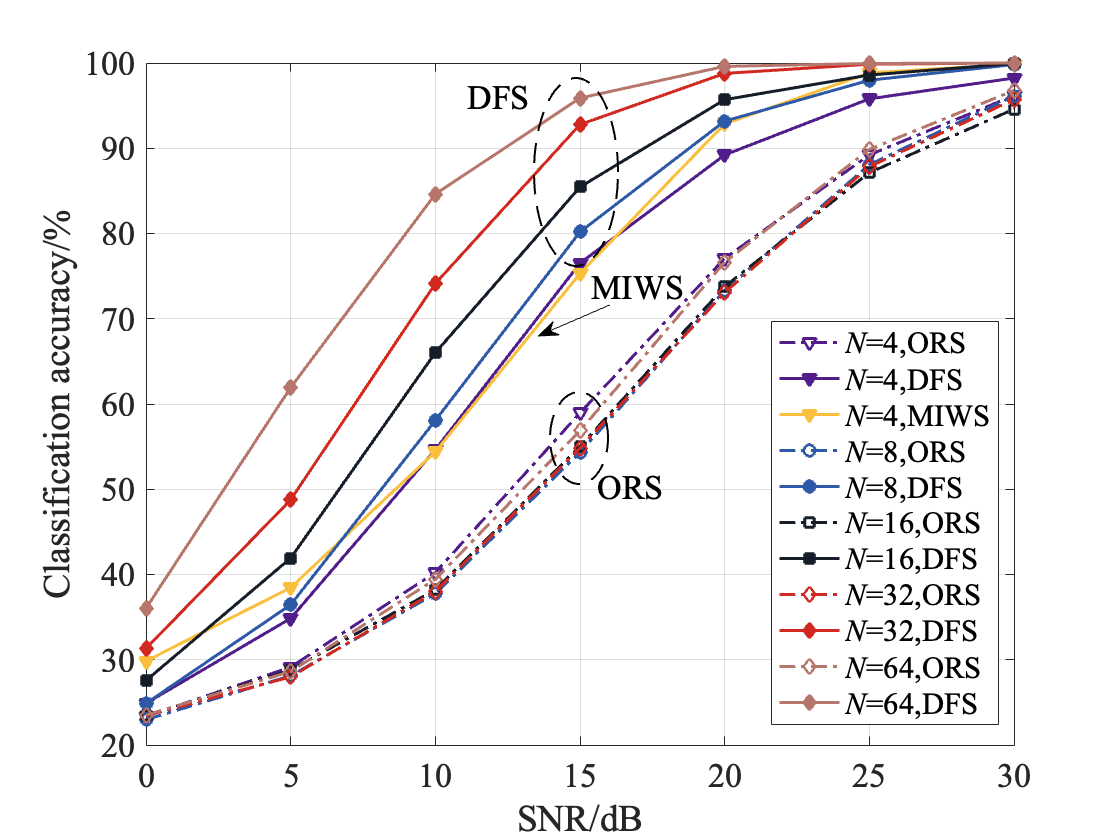}}
  \caption{\centering{Classification accuracy of DFS based on LMS with $M=5$, and $\chi=0.01$.}}
  \label{DFS_4}
\end{figure}
Fig.~\ref{DFS_4} shows the RFFI results of DFS, where the LMS-based feature extraction method is adopted with $M=5$ and $\chi=0.01$ for all receiver antennas.
As can be seen from this figure, the superiority of DFS over ORS becomes increasingly apparent as $N$ increases.
Such phenomenon that DFS outperforms ORS in terms of RFFI accuracy demonstrates that DFS can effectively filter out channel noise and receiver distortions.
Moreover, there are two noteworthy points: 1) when $N=4$, DFS performs worse than MIWS, suggesting that MIWS is more appropriate when $N\le4$; 
2) when SNR$=$0dB, the performance gain of DFS over ORS is not obvious, whereas when SNR$>$15dB, such gain becomes more significant for different $N$s, indicating that the level of the performance gain of DFS with respect to ORS is related to the SNR. These two observations are consistent with the statements in Section IV.

\begin{figure}[htbp]
  \centerline{\includegraphics[width=0.5\textwidth]{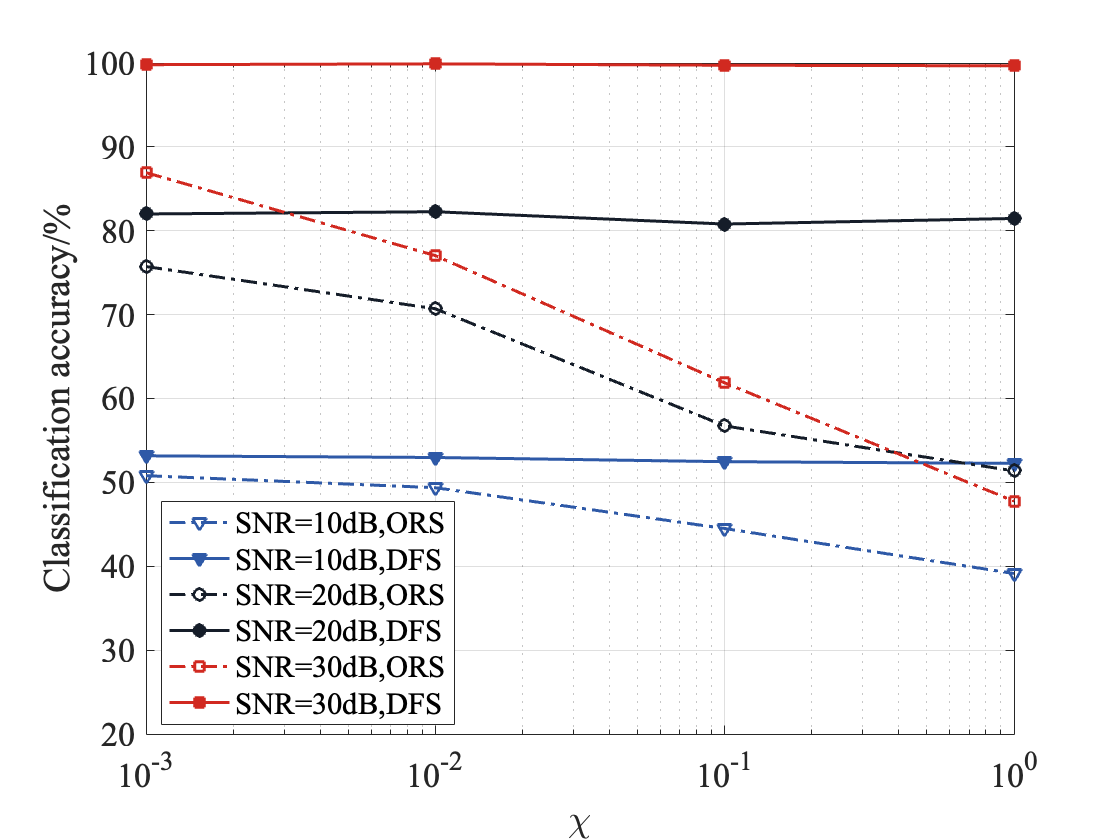}}
  \caption{\centering{Classification accuracy of DFS based on ITD with $M=5$, and $N=8$.}}
  \label{DFS_2}
\end{figure}
\begin{figure}[htbp]
  \centerline{\includegraphics[width=0.5\textwidth]{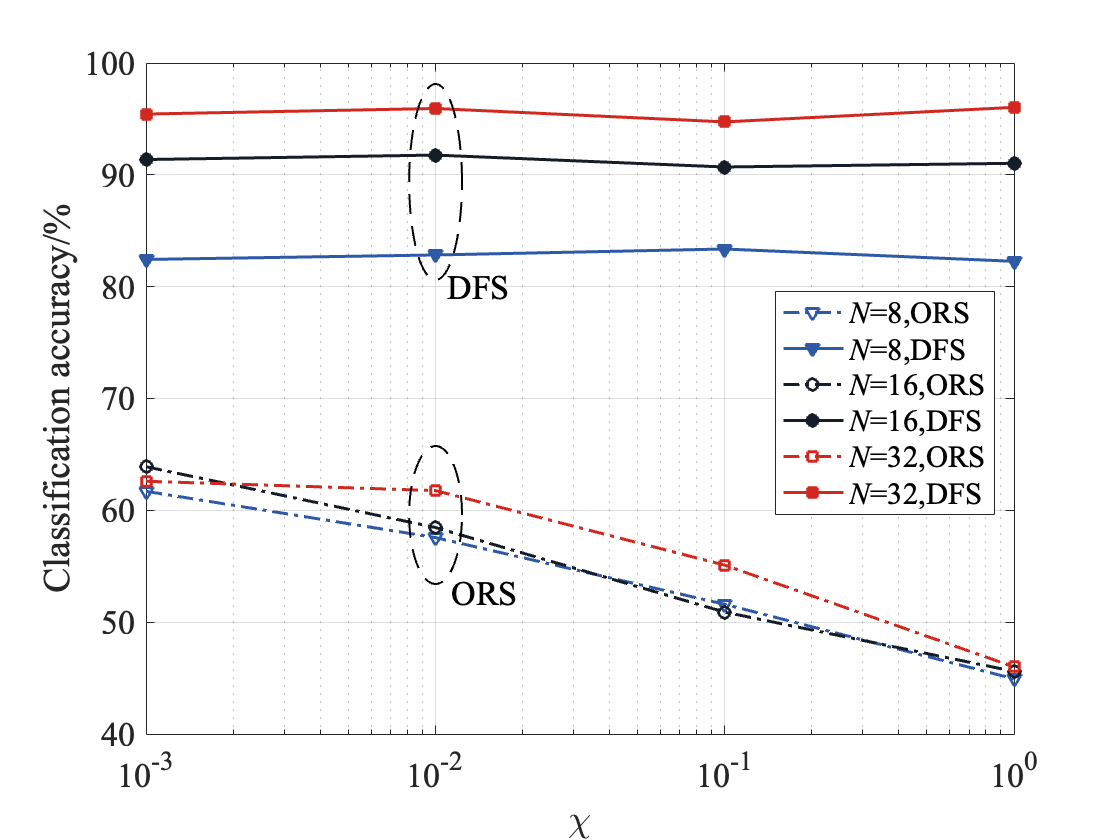}}
  \caption{\centering{Classification accuracy of DFS based on ITD with $M=5$, and SNR=15dB.}}
  \label{DFS_3}
\end{figure}
Fig.~\ref{DFS_2} and Fig.~\ref{DFS_3} show the RFFI accuracy versus receiving phase noise when using the ITD-based RF feature extraction method. 
The former varies SNR with $N=8$, and the latter varies $N$ with SNR$=15$dB.
As seen from these two figures, the larger the $\chi$, the worse the RFFI accuracy of ORS. In contrast, DFS shows its stability and robustness under different $\chi$, which indicates that DFS filters out receiver phase noise effectively. 
Fig.~\ref{DFS_3}  illustrates that the performance of DFS becomes better with the increasing number of receiving antennas, which remains consistent with the conclusion of Fig.~\ref{DFS_4}.
\subsection{Experimental results of GDFWS}
\begin{figure}[htbp]
  \centerline{\includegraphics[width=0.5\textwidth]{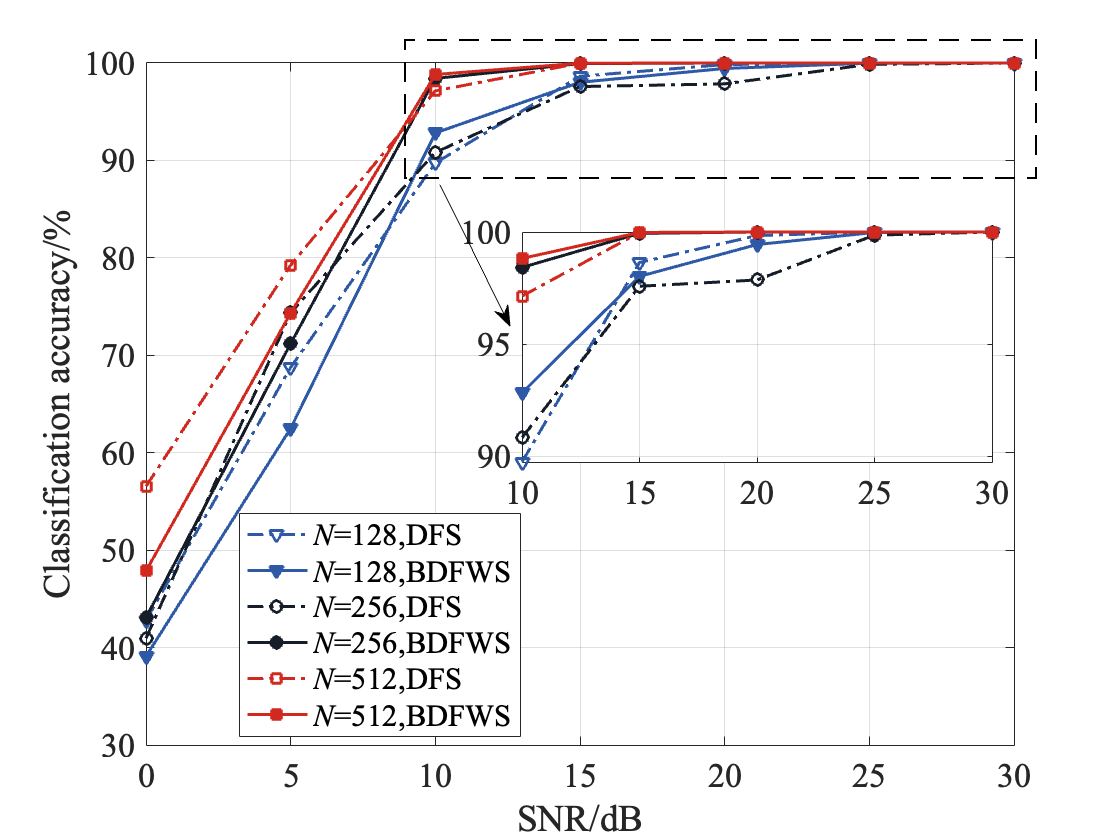}}
  \caption{\centering{Classification accuracy of GDFWS based on LMS with $M=5$, and $\chi=0.01$.}}
  \label{BDFWS_2}
\end{figure}
Figure~\ref{BDFWS_2} provides the performance of BDFWS in comparison with DFS when the LMS-based RF feature extraction method is employed. 
The receiving antennas are evenly divided into four groups, and the other settings are consistent with those in Fig.~\ref{DFS_4}.
When SNR$\ge$10 dB and $N=256$ or $N=512$, GDFWS with weighted voting outperforms DFS. 
However, when SNR$<$10 dB, performance saturation does not appear in DFS, thus its performance is comparable with that of GDFWS.
Moreover, when $N=128$, the overall performance of GDFWS is inferior to that of DFS, which suggests that GDFWS is unsuitable for scenarios where $N\le128$. 
Overall, the results presented in this figure demonstrate the effectiveness of GDFWS with weighted voting in scenarios where $N>128$ and SNR$\ge$10 dB when compared with DFS, the finding of which is consistent with the theoretical analysis presented in Section IV.
\section{Conclusion}
This paper investigates three RFFI schemes to cater to the different numbers of receiving antennas.
When the number is small, we propose MIWS that uses the weighted voting of intermediate classification results for RFFI. 
For a moderate quantity of receiving antennas, DFS is proposed to perform statistical averaging to filter out channel noise and receiver distortions. 
If a large amount of receiving antennas are available, GDFWS, which enjoys the advantages of both MIWS and DFS,
is developed to solve the performance saturation problem in DFS and improve classification accuracy.
We further study the impact of the number of receiving antennas on DFS performance, and provide guidelines on selecting appropriate schemes for different scenarios in the following aspects: 
1) When the number of antennas is $N \le 4$, MIWS is recommended. 
2) When the number of antennas is $4<N\le128$, DFS is the best choice.
3) The performance saturation in DFS occurs commonly when SNR is high. Hence, when $N > 128$ and SNR$\ge$10dB, GDFWS is preferable.

\bibliographystyle{ieeetr}
\bibliography{reference} 
\end{document}